\documentclass[showpacs, twocolumn]{revtex4-1}
\usepackage{graphicx}
\usepackage{amsmath}
\usepackage{csquotes}

\begin{document}

\title{ Singular Soliton Molecules of the Nonlinear Schr\"odinger Equation }

\author{Khelifa Mohammed Elhadj$^{1,2}$, L. Al Sakkaf$^{3}$,\,U. Al Khawaja$^{3}$ and Abdel\^{a}ali Boudjem\^{a}a$^{1,2}$ }

\affiliation{$^1$Department of Physics, Faculty of Exact Sciences and Informatics, Hassiba Benbouali University of Chlef, P.O. Box 78, 02000, Chlef, Algeria.\\
$^2$Laboratory of Mechanics and Energy, Hassiba Benbouali University of Chlef, P.O. Box 78, 02000, Chlef, Algeria.
\\$^3$Department of Physics, United Arab Emirates University, P.O. Box 15551, Al-Ain, United Arab Emirates.} 



\begin{abstract}
We derive an  exact solution to the local nonlinear Schr\"odinger equation (NLSE) using the Darboux transformation method. 
The new solution describes the profile and dynamics of a two-soliton molecule. Using an  algebraically-decaying seed solution, 
we obtain a two-soliton solution with diverging peaks, which we denote as singular soliton molecule.
We find that the new solution has a finite binding energy. We  calculate the force  and potential of interaction between the two solitons, 
which turn out to be of  molecular-type.
The robustness of the bond between the two solitons is also verified. Furthermore, we obtain a new solution to the nonlocal NLSE using the same method and  seed solution. 
The new  solution in this case  corresponds to  an elastic collision  of  a soliton,  a breather soliton on flat background, and a breather soliton  on a background with linear ramp. 
Finally, we consider an NLSE which is nonlocal in time rather than space. Although we did not find a Lax pair to this equation, we derive three exact solutions.

\end{abstract}

\maketitle

\section{Introduction} 

The fundamental NLSE is one of the most investegated equations  in describing the dynamics of multiple physical phenomena, in both discrete and continuous systems. 
It describes Bose-Einstein condensates  (alias Gross-Pitaevskii equation) \cite{Bong}, the collapse of plasma waves, pulses in nonlinear optical fibers  \cite{Hasegawa}, 
the propagation of waves in nonlinear waveguides, and the interaction between solitons in nonlinear waveguides. 
One important admitted solution to this equation is soliton. It is a  localized wave  originating from the competition between dispersive and nonlinear effects \cite{Rem}. 
In fact, it is the most essential phenomenon of the local NLSE. 
Solitons appear in many diverse systems such as plasmas, astrophysics, molecular biology, nonlinear optics, spin waves, superfluidity, and Bose-Einstein condensates 
(see for review Ref. \cite{Kiv}).

Many successful techniques have been developed to solve  the NLSE.  Among them are Painlev\'e analysis  \cite{Conte,Usama,Brug},
Hirota method \cite{Hiro}, similarity transformation method \cite {Dai}, Lax Pair (LP) and Darboux transformation (DT) method \cite{Lax, DT, DT1},
Miura transformation \cite{Miur,Miur1}, inverse scattering transform and Hamiltonian approach \cite {Ablz}, 
homotopy analysis method \cite{LL}, Exp-function method \cite{Ganj}, the tanh-function method \cite{Abdu,Yan},
the homogeneous balance method \cite{Wang},  and the F-expansion method \cite{Zhou}.

Currently, there is a considerable interest in finding exact solutions to the nonlocal NLSE \cite{Ab,Ab1,sa,kh, vi}. 
This  non-Hermitian and  PT-symmetric equation with the potential $V(x,t)=u(x, t) u^*(-x, t)$, where $u(x,t)$ is the meanfield wavefunction, 
satisfies the PT-symmetric condition, $V(x,t)= V^*(-x,t)$.   
Several efforts were devoted in showing that this equation admits a soliton solution as well \cite{Ab,Ab1,sa,kh, vi}. 

Briefly, the  LP and DT method is based on  searching for an appropriate pair of matrices that associates the nonlinear system to a linear system.  
The LP should be associated with the nonlinear model through what is called a compatibility condition. 
The obtained linear system is  solved using a seed solution, $u_0(x,t)$, which is a known exact solution of the nonlinear system. 
Each seed solution generates a family of exact solutions.   

Prominent among the solutions of the NLSE is the two-soliton solution which can be obtained using the LP and DT method  \cite {usaboudj, khaw}. 
Employing the trivial  seed solution, $u_0(x,t)=0$,  the DT generates a single soliton solution. Using the latter as seed,  generates the two-soliton solution \cite{khaw}. 
The binding energy, the force, and potential of interaction between solitons have been calculated and studied extensively \cite{Malom, serhasg,khaw, usaboudj, Mitch10, Turi, abdou, abdou3}.

The interest in two-soliton solution stems, not only from its importance on fundamental level, but also from its tremendous application  
as a data carrier in optical fibers \cite{Mitch10} . It has been suggested that such a soliton-molecule may increase the data-carrying capacity \cite{Hasegawa, Mitch05, Mitch12}.
The existence of a nonzero binding energy in terms of the width of the soliton is a signature on its stability. 

Here, we follow the above-mentioned effort in the literature to find new exact solutions to the NLSE.
Specifically, we use a seed solution of the form $u_0(x,t)=1/x$ to generate a two-soliton solution that is characterized by two diverging peaks known as singular solitons. 
It is found that such states are generated due to the strong self-repulsion \cite{Veron, Hid}.
Despite its divergency, the new solution describes a soliton molecule with a binding energy. 
We have calculated the potential of the interaction between the two solitons to show that it is of molecular type. 
To the best of our knowledge, this is a new kind of soliton molecule.

Furthermore, we have considered the nonlocal NLSE and used the same procedure to generate a new solution out of the $u_0(x,t)=1/x$ seed. 
It turned out that the new solution is much richer than the local case. 
Here, the new solution corresponds to the scattering of stationary soliton and two breathers on a finite background at half of space and inclined background at the other half.
 
The plan of this work is as follows.
In Sec.\ref{model}, we derive a new two-soliton solution with two diverging peaks to the local NLSE using the LP and DT method with an algebraically-decaying seed. 
We then investigate  the binding energy, the force,  and potential of interaction for these new local NLSE solution, to check that it is indeed a soliton molecule.
We also consider its scattering properties with other solitons to show that the integrity of individual solitons as well as the molecule is preserved after scattering.
In Sec.\ref{model2}, we employ the same seed to obtain a new exact solution to the nonlocal NLSE, which demonstrates an elastic interaction between one bright soliton 
and two breather solutions, one on a finite flat background and the other on a ramp background. 
Section \ref{model3} is devoted to the case with reverse-time nonlocal NLSE.
We end with a summary of our main conclusions in Sec.\ref{Conc}.

\section{New Exact Solution to the local NLSE } \label{model}

In this section, we apply the LP and DT method with the rational seed solution,   $u_0 (x,t)=1/x$, to the local NLSE given by 
\begin{equation} \label{NLSE}
i u_t+\frac{1}{2} u_{xx}-|u|^2 u=0,
\end{equation}
The Lax pair of Eq. (\ref{NLSE}) is given by \cite{DT1}
 \begin{equation} \label{pot2}
 \Phi _x=J\cdot\Phi\cdot\Lambda +U\cdot\Phi,
 \end{equation}
 and
 \begin{equation} \label{pot3}
 \Phi _t=i \,J\cdot\Phi\cdot\Lambda^2+i\,U\cdot\Phi\cdot\Lambda +V\cdot\Phi.
 \end{equation}
 The matrices $U$, $V$, $\Lambda$, and $J$ are defined as
 \begin{equation} \label{pot4}
 U=\left(
 \begin{array}{cc}
 0 & u \\
 -u^*& 0 \\
 \end{array}
 \right),
 \end{equation}
\begin{equation} \label{pot5}
V= \frac{i}{2} \left(
\begin{array}{cc}
\left| u\right| ^2 &  u_x \\
u_x^*& - \left| u\right| ^2 \\
\end{array}
\right),
\end{equation}
\begin{equation} \label{pot6}
\Lambda =\left(
\begin{array}{cc}
\lambda _1 & 0 \\
0 & \lambda _2 \\
\end{array}
\right),
\end{equation}
\begin{equation} \label{pot7}
J=\left(
\begin{array}{cc}
	1 & 0 \\
	0 & -1 \\
\end{array}
\right),
\end{equation}
where $u^*(x,t)$ is the complex conjugate of $u(x,t)$ and $\lambda _{1,2}=\lambda _{(1,2)r}+i \lambda _{(1,2)i}$  
are the spectral complex parameters  with $\lambda _{(1,2)r}$ and $\lambda _{(1,2)i}$ are arbitrary real constants. The auxiliary field $\Phi(x,t)$ is given by
\begin{equation} \label{pot1}
\Phi (x,t)=\left(
\begin{array}{cc}
\psi _1(x,t) & \psi _2(x,t) \\
\phi _1(x,t) & \phi _2(x,t) \\
\end{array}
\right).
\end{equation}
The compatibility condition $\phi _{xt}=\phi _{tx} $ leads to
\begin{equation} \label{pot8}
U_t-V_x+ \bigg[U,V \bigg ]=0,
\end{equation}
where $[U,  V]$ is the commutator between $U$ and $V$. 
The use of Eqs. (\ref{pot4}), (\ref{pot5}) and (\ref{pot8}) yields the compatibility condition which establishes the link between the NLSE and the LP. 
The DT is defined by \cite{DT1}
\begin{equation} \label{pot9}
\Phi [1]=\Phi \cdot \Lambda - \sigma\,\Phi,
\end{equation}
where $\Phi[1]$ is the transformed field, and $[J,\sigma]$ is the commutator between $J$ and $\sigma$, with $\sigma$  given by
\begin{equation} \label{pot10}
\sigma=\Phi _0 \cdot \Lambda \cdot \Phi _0^{-1}.
\end{equation}
Here, $\Phi_{0}$ is a seed solution of the linear system for a given seed solution of the NLSE, $u_0(x,t)$.  
The field $\Phi $ represents any solution of the linear system and  $\Phi [1] $ is the new solution of this system which obeys
\begin{equation} \label{pot11}
\Phi[1]_x=J\cdot\Phi [1]\cdot\Lambda+U[1]\cdot\Phi [1],
\end{equation}
\begin{equation} \label{pot12}
\Phi[1]_t=i \,J\cdot\Phi [1]\cdot\Lambda^2+i\,U[1]\cdot\Phi [1]\cdot\Lambda +V[1]\cdot\Phi [1],
\end{equation}
where
\begin{equation} \label{pot13}
 U[1]= U_0+ \bigg[J, \sigma \bigg],
\end{equation}
\begin{equation} \label{pot14}
V[1]=V_0+\bigg[U_0,\sigma \bigg],
\end{equation}
and $U_0$ and $V_0$ are the LP in terms of the seed solution. 
The matrices $J$ and $\Lambda$ are constant and do not change under the DT. 
Equations (\ref{pot4}), (\ref{pot5}) and (\ref{pot9}) give the new solution of the NLSE as follows 
\begin{equation} \label{pot15}
 U [1]=U_0+Q\left(\psi _1,\phi _1\right), 
\end{equation}
where the Darboux dressing is given by
\begin{equation} \label{dressing}
Q\left(\psi _1,\phi _1\right)= [2 \left(\lambda _1-\lambda _2\right) \phi _1 \psi _1]/(\phi _1 \psi_2-\phi _2 \psi _1).
\end{equation}
Using the following symmetry reductions
\begin{equation} \label{syred} 
\lambda _2{}^*=-\lambda _1, \qquad \phi _2{}^*=\psi _1, \qquad \psi _2{}^*=\phi _1,
\end{equation}
reduces Eqs. (\ref{pot2}) and (\ref{pot3}) for  $\psi _1$ ,$\phi_1$, $\psi _2$, and $\phi_2$,  to
\begin{equation} \label{pot16}
-\lambda _1 \psi _1-i u \phi _1+\psi _{1x}=0,
\end{equation}
\begin{equation} \label{pot17}
\lambda _1 \phi _1+i u^* \psi _1+\phi _{1x}=0,
\end{equation}
\begin{equation} \label{pot18}
i \psi _{1t}+\psi _1 \left(\lambda _1^2+\frac{\left| u\right| ^2}{2}\right)+\phi
_1 \left(\lambda _1 u+\frac{u_x}{2}\right)=0,
\end{equation}
\begin{equation} \label{pot19}
i \phi _{1t}+\psi _1 \left(\frac{\left(u_x\right)^*}{2}-\lambda _1
u^*\right)-\phi _1 \left(\lambda _1^2+\frac{\left| u\right|
	^2}{2}\right)=0.
\end{equation}
Substituting  the seed $ u_0=1/x$ in (\ref{pot16})-(\ref{pot19}) and solving for $\psi_1$ and $\phi_1$ reads
\begin{equation} \label{pot}
\psi_1(x,t)=\frac{e^{-x \lambda_1} }{4x}\left[4 \, c_{1}\,e^{-i \lambda_1^2 t}+\frac{c_{2}(2\,\lambda_1 \,x-1)e^{2\lambda_1 x+i\lambda_1^2 t}}{\lambda_1^2}\right],
\end{equation}
\begin{equation} \label{pot20}
\phi_1(x,t)=\frac{i \,e^{-\lambda_1(x+i \lambda_1 t)} }{4 \,\lambda_1^2 \,x}\left[4 \,c_{1} \lambda_1^2 (2 \,\lambda_1 x+1)-c_{2} \,e^{2 \lambda_1(x+i \lambda_1 t)}\right],
\end{equation}
where $\ c_{1}$ and $\ c_{2}$ are arbitrary complex constants. Substituting (\ref{pot}) and (\ref{pot20}) into (\ref{pot15}), 
we obtain a new exact solution to the local NLSE, Eq. (\ref{NLSE}) namely:
    \begin{widetext}    	
\begin{equation} \label{pot21}
    u_1(x,t)=\frac{16\,A_1\,x_1^2+2\,A_1\,[2\,A_1\, x_0\, (\lambda_1 -\,\lambda_1^*+2\,x_1\, x)-\,A_0 \,\lambda_1^2\, e^{2\, \lambda_1^*(\,x-\,i\,\lambda_1^*\, t)}]\,e^{2\,\lambda_1 (\,x+\,i \,\lambda_1\, t)}}{4 \,A_0\,\lambda_1^2 \,x_0 \,e^{2\,\lambda_1^* (\,x-\,i \,\lambda_1^*\,t)}+\,A_1\,[2\, A_1\, x_0\,+A_0\,(2\,x_1\, x\, -\,x_0)\,e^{2\,\lambda_1^*(\,x-\,i\,\lambda_1^*\, t)}]\,e^{2\,\lambda_1(\,x+\,i\,\lambda_1\, t)}-8\, A_1 \,\lambda_1^2\,(\,x_0 +2\, x_1\, x)},
\end{equation}   	
    \end{widetext}
where $x_0={\lambda_1}+{\lambda_1^*}$, $x_1={\lambda _1}{\lambda_1^*}$, $A_0= c_{2} \,c_{2}^*/2 {\lambda_1^*}^2 c_{1}\, c_{1}^* $, and $A_1=c_{2}/c_{1}$.\\

\begin{figure} 
	\centerline{
		\includegraphics[width=3 cm,height=4 cm]{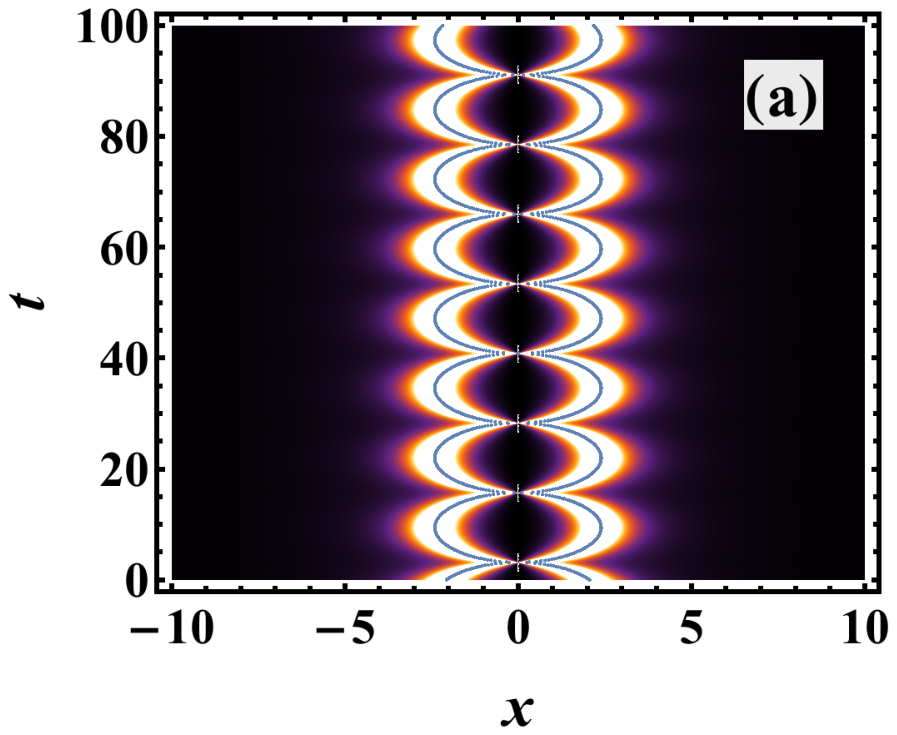}
		\includegraphics[width=3. cm,height=4 cm]{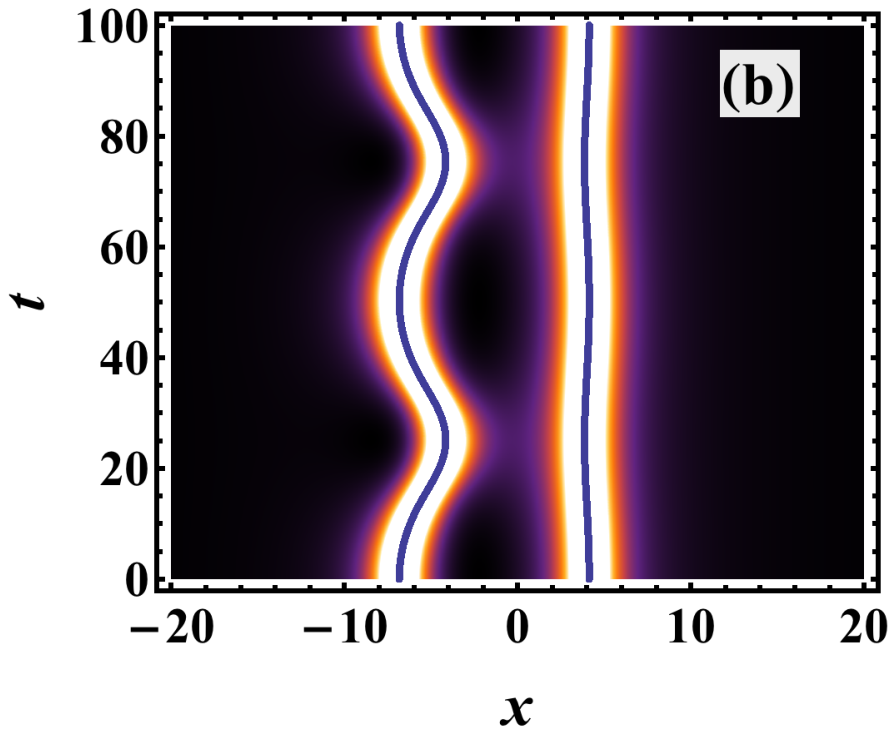}
		\includegraphics[width=3. cm,height=4 cm]{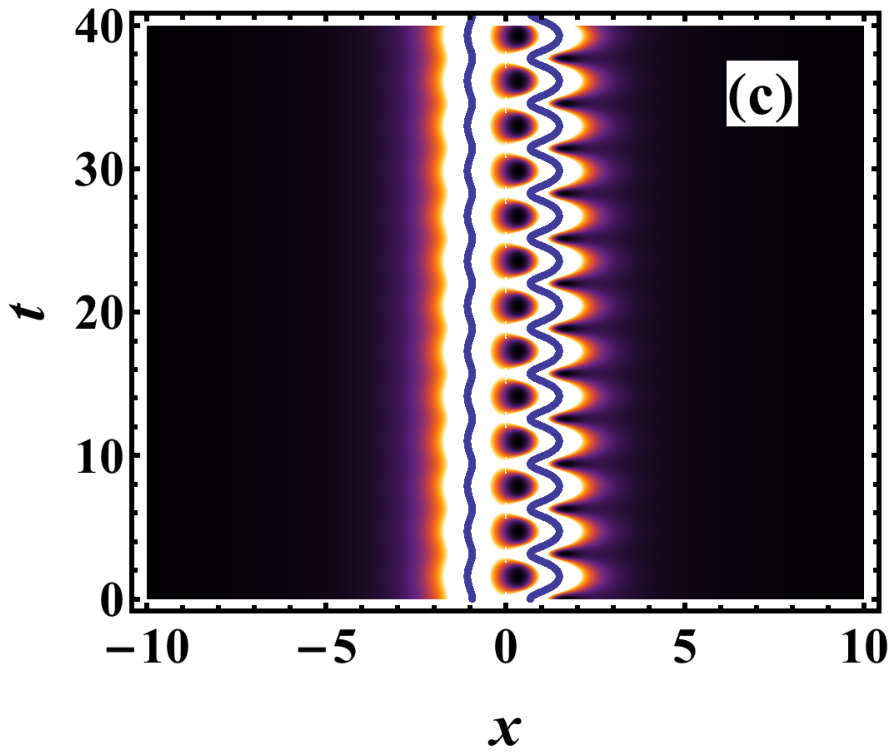}}
	\caption{ Singular solitons molecules with zero relative velocity. (a)  Symmetric coalescing solitons (\ref{pot21mol}).
 (b) Asymmetric noncoalescing solitons (\ref{pot22}). (c) Asymmetric coalescing solitons (\ref{pot23}).
Blue lines show the soliton's trajectory.}
	\label{solm}
\end{figure}

Depending on the values of these parameters, one can distinguish three different regimes, namely;
\begin{flushleft}
	\bf
(i) Symmetric coalescing solitons: 
\end{flushleft}
For $ \lambda_1=\lambda_{1r}=1/2$, $c_1=10+10 i$ and $c_2=10-10 i$, the solution (\ref{pot21}) reduces to 
\begin{equation}\label{pot21mol}
\ u_{1}(x,t)=\frac{e^{\frac{i t}{2}} \left(2 x e^{x+\frac{i t}{2}}-i e^{2 x}+i\right)}{i
	(x-2) e^{2 x+\frac{i t}{2}}+2 e^{x+i t}-i e^{\frac{i t}{2}} (x+2)-2 e^x},
\end{equation}
which is displayed in Fig. \ref{solm}(a). Here, we observe that the two solitons collide periodically. 
\begin{flushleft}
	\bf
(ii) Asymmetric noncoalescing solitons:
\end{flushleft}
 For $\lambda_1=1/4$, $c_1=2i$ and $c_2=-4 i$, we get
\begin{equation}\label{pot22}
u_{1}(x,t)=\frac{e^{\frac{i t}{8}} \left(8 x e^{\frac{x}{2}+\frac{i t}{8}}+64
	e^x-1\right)}{-64 (x-4) e^{x+\frac{i t}{8}}+32 e^{\frac{x}{2}+\frac{i
			t}{4}}+e^{\frac{i t}{8}} (x+4)+32 e^{x/2}}.
\end{equation}
Here the two solitons still  form a molecule but they do not coalesce, as is seen in Fig. \ref{solm}(b).

\begin{flushleft}
	\bf
	(iii) Asymmetric coalescing solitons: 
\end{flushleft}
 For small intersoliton distance,  i.e.,  $\lambda_1=1$, $c_1= 1+i$ and $c_2=1+i$, Eq. (\ref{pot20}) simplifies to
\begin{equation} \label{pot23}
 \ u_{1}(x,t)=\frac{e^{2 i t} \left(16 x e^{2 x+2 i t}-e^{4 x}+16\right)}{(x-1) e^{4 x+2 i
 		t}+4 e^{2 x+4 i t}-16 e^{2 i t} (x+1)+4 e^{2 x}}.
 \end{equation}
In this case the solitons coalesce,  as in the first case, but the inter-soliton oscillation is performed mainly by one soliton, as  in the second case.  
This is shown in Fig. \ref{solm}(c).

\subsection{Interactions of singular solitons} \label {ISS}

\begin{figure} 
	\includegraphics[scale=0.6]{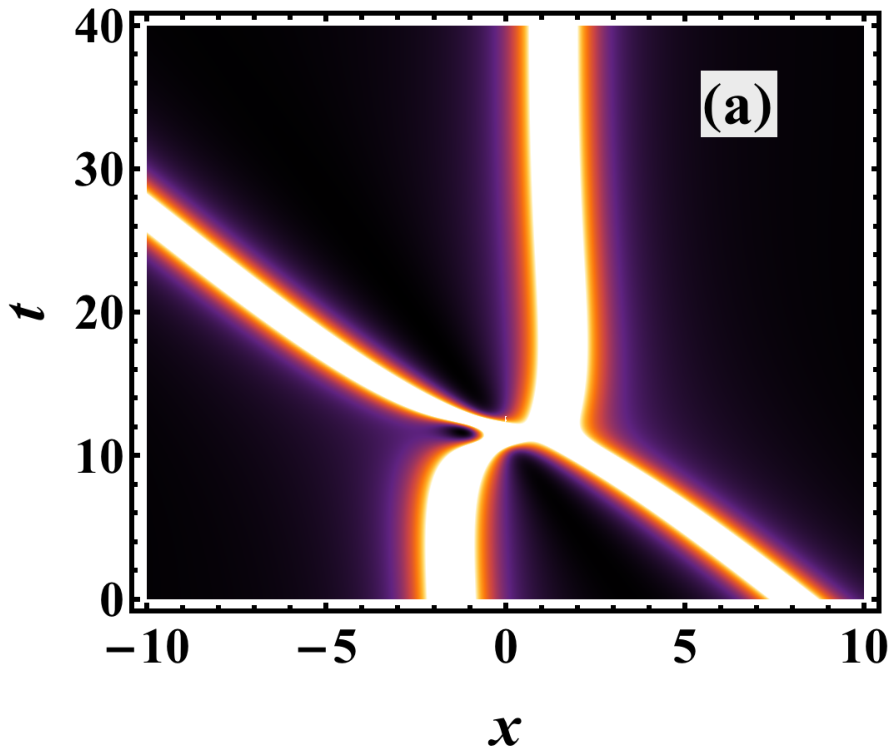}
	\includegraphics[scale=0.6]{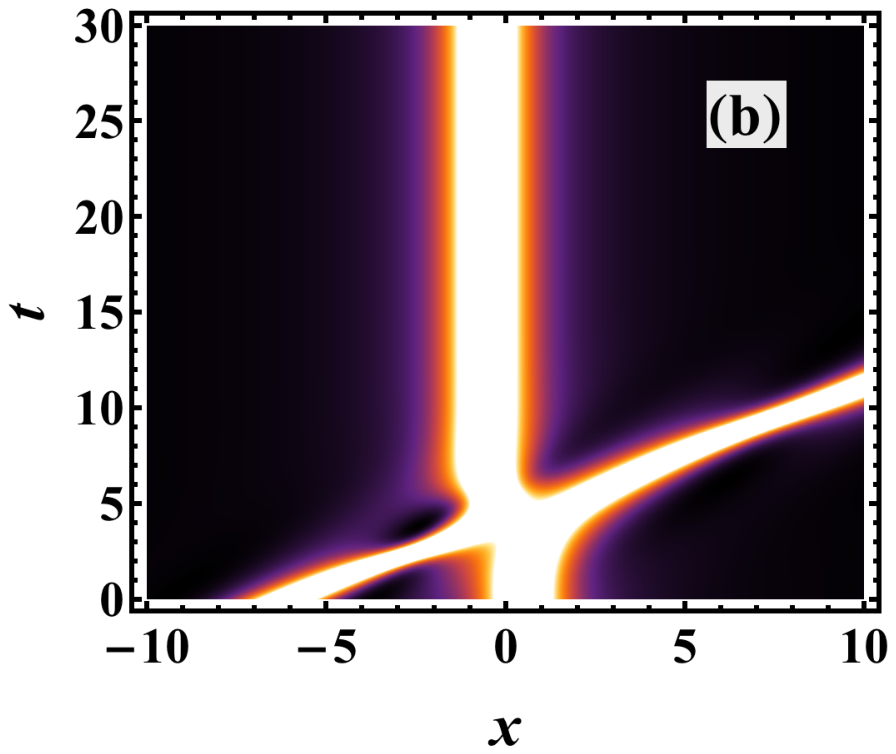}
	\caption{Transmission of one soliton with speed  $\lambda_{1i}$  through a stationary soliton. 
       Parameters are: (a) $\lambda_1= -0.15  - 0.3 i$, $c_1= -10-10 i$, and $c_2= 10+50 i$. 
                                (b) $\lambda_1= -0.3 + 0.7 i$, $c_1= 10+ 10 i$, and  $c_2= -1$.} 
	\label {trans}
\end{figure}

Let us now discuss the interaction of the two local solitons given in  Eq. (\ref{pot21}). 
Figures \ref{trans}(a) and \ref{trans}(b) show the scattering of  two solitons where they preserve their integrity after collision.

To determine the force of interaction between the two solitons, we calculate first the acceleration, $a_r$ of the right soliton and $a_l$ of the left soliton. 
This is performed by extracting the center of mass of
each soliton, $x_r$ for the right soliton and $x_l$ for the left soliton, and then taking the second derivative, $a_r=\ddot {x_r}$ and $a_l=\ddot {x_l}$. 
The force of interaction between the two solitons is proportional to the second derivative of their separation $F=m (a_r-a_l)$,  where
$m=\int_{-\infty}^{\infty} |u_1(x,t)|^2dx$. Then, the integration of the force gives the potential $V=-\,\int F dx$.

When the two solitons are well-separated from each other, analytic expressions for their positions can be extracted from the exact solution,
Eq. (\ref{pot21}), which lead to the following expressions for their accelerations: 
	\begin{equation}\label{pot24}
	\ a_{r}=\frac{-x^3+4 x^2-13 x+64 e^x (4-x)-4}{8 e^x \left(x^3-6 x^2+12 x-8\right)+1536
		\left(x^3-2 x^2-4 x+8\right)},
	\end{equation}
	and 
	\begin{equation}\label{pot25}
	\ a_{l}=\frac{64 \left(x^3-4 x^2+13 x+4\right)+e^x (x-4)}{512 e^x \left(x^3-6 x^2+12
		x-8\right)+48 \left(x^3-2 x^2-4 x+8\right)}.
	\end{equation}
	
Similarly, when the solitons are close to each other, the accelerations read
	\begin{equation} \label{pot26}
	\ a_{r}=\frac{64 \left(-16 x^3-16 x^2-13 x+1\right)+4 e^{-x} (-x-1)}{-384 x^3-192
		x^2+e^{-4 x} \left(-8 x^3-12 x^2-6 x-1\right)+96 x+48},
	\end{equation}
	and
	\begin{equation}\label{pot27}
	\ a_{l}=\frac{4 \left(-16 x^3-16 x^2-13 x+1\right)+64 e^{-4 x} (-x-1)}{16 e^{-4 x}
		\left(-8 x^3-12 x^2-6 x-1\right)+3 \left(-8 x^3-4 x^2+2 x+1\right)}.
	\end{equation}

\begin{figure}
 	\centerline{
  		\includegraphics[width=4.3  cm,height=4.3cm]{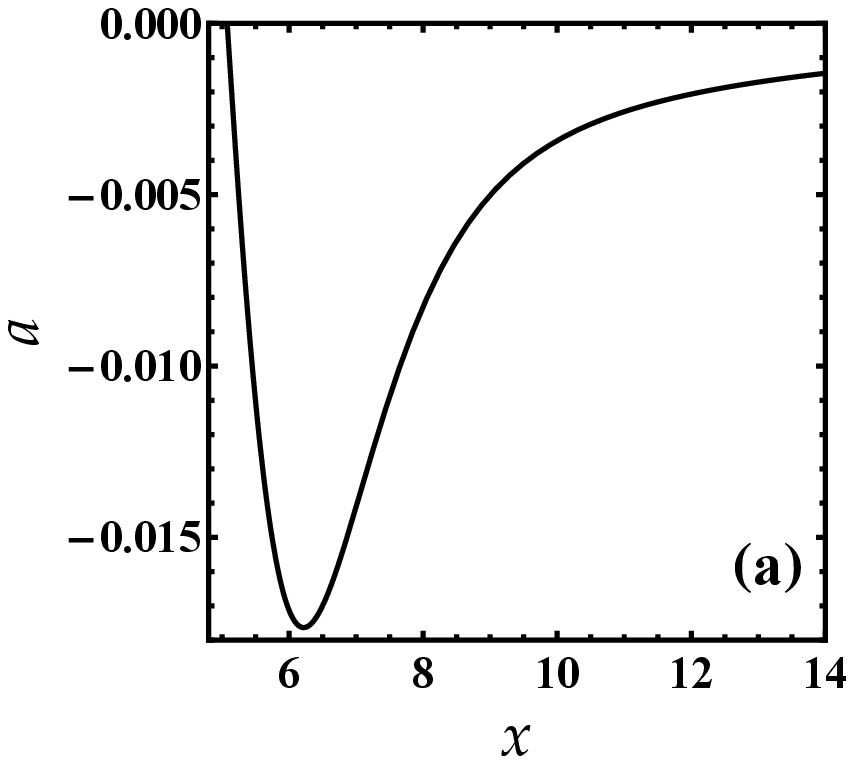}
 		\includegraphics[width=4.3 cm,height=4.3cm]{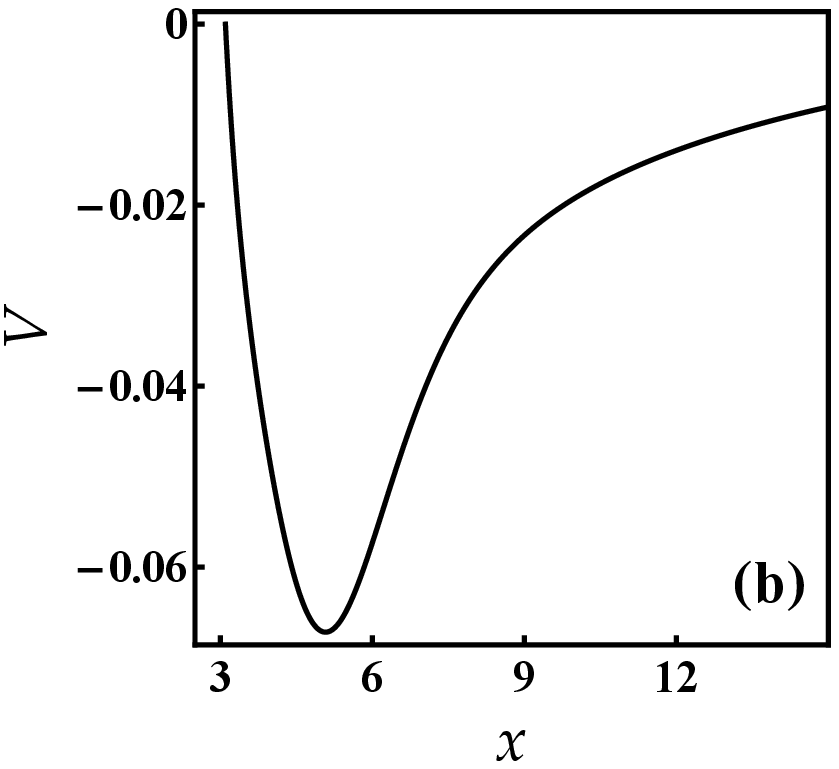}}
 	\caption{The force and potential of interaction when the solitons are well-separated. (a) The acceleration of the solitons separation, $a_r-a_l$, 
as given by  Eqs. (\ref{pot24}) and (\ref{pot25}),  which is proportional to the mutual force of interaction, $F$. (b) Interaction potential energy.
Parameters are : $\lambda_1=1/4$, $c_1=2i$, and $c_2=-4 i$.}
\label {separ}
 \end{figure}

\begin{figure}
	\centerline{
		\includegraphics[width=4.3 cm,height=4.3 cm]{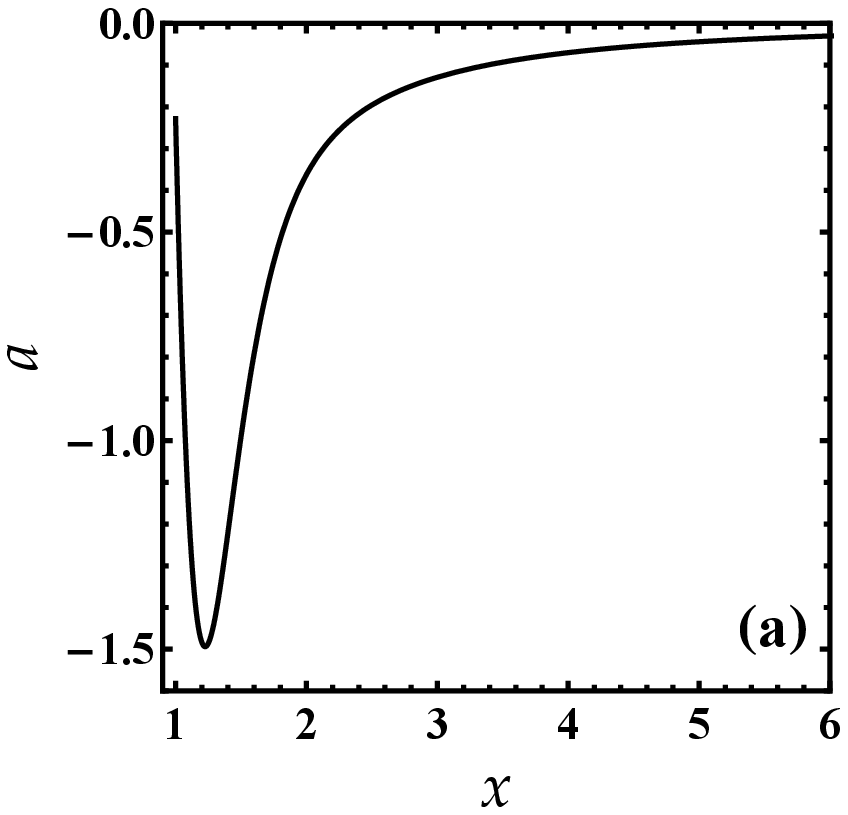}
		\includegraphics[width=4.3 cm,height=4.3 cm]{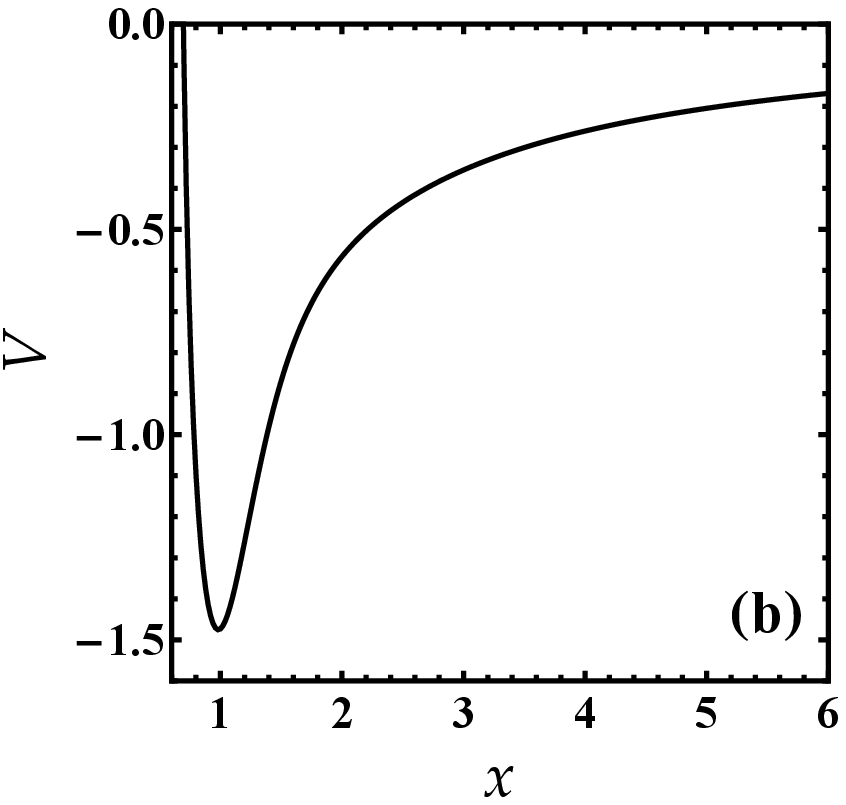}}
 	\caption{The force and potential of interaction when the solitons are closely-separated. 
(a) The acceleration of the solitons separation, $a_r-a_l$, as given by  Eqs. (\ref{pot26}) and (\ref{pot27})  which is proportional to the mutual force of interaction, $F$. 
(b) Interaction potential energy. Parameters are : $\lambda_1=1$, $c_1= 1+i$, and $c_2=1+i$.}
\label {Fus}
\end{figure}

It is clearly apparent from Fig. \ref{separ}(b) that for large separation, the interaction potential is negative 
showing the stability and the robustness of the bond between the solitons. 
It is noticed that while each of the two solitons experiences the same force of interaction, 
they have different acceleration due to their different masses, as is shown in Figs. \ref{separ}(a) and \ref{solm}(b).   
The situation is quite different when the two solitons are placed close to each other; the acceleration and the interaction potential well become narrower and deeper 
a fact that influences the formation of the bound state as is seen in Figs. \ref{Fus}(a) and \ref{Fus}(b), respectively.
Figures \ref{separ} and \ref{Fus} clearly show molecular type of  potential  between the two singular solitons.

\section{New Exact Solution to the Nonlocal NLSE } \label{model2}

In this section, we apply the LP and DT method with the same rational seed solution,  $u_0(x,t)=1/x$, to the nonlocal NLSE 
which can be written as:
\begin{equation} \label{nonNLSE}
i u_t+\frac{1}{2} u_{xx}+u^2 \bar{u}=0,
\end{equation}
where $\bar{u}=u^*(-x,t)$. 
The Lax pair of Eq. (\ref{nonNLSE}) can be found via Eqs. (\ref{pot2}) and (\ref{pot3}) with the following operators
\begin{equation} \label{pot4non}
U=\left(
\begin{array}{cc}
0 & -u \\
-\bar{u} & 0 \\
\end{array}
\right),
\end{equation}
\begin{equation} \label{pot5non}
V= \frac{i}{2} \left(
\begin{array}{cc}
 u\,\bar{u}&  u_x \\
\bar{u} & -u\,\bar{u}\\
\end{array}
\right),
\end{equation}
\begin{equation} \label{pot6non}
\Lambda =\left(
\begin{array}{cc}
\lambda _1 & 0 \\
0 & \lambda _2 \\
\end{array}
\right),
\end{equation}

\begin{equation} \label{pot7non}
J=\left(
\begin{array}{cc}
1 & 0 \\
0 & -1 \\
\end{array}
\right).
\end{equation}
With the use of the symmetry reductions in (\ref{syred}),  the linear system, (\ref{pot2}) and (\ref{pot3}), is reduced to 
\begin{equation}\label{non1}
u\,\phi_1-\lambda_1\,\psi_1+{\psi_1}_x=0,
\end{equation}
\begin{equation}
\bar{u}\,\psi_1-\lambda_1\,\phi_1-{\phi_1}_x=0,
\end{equation}
\begin{equation}
-i\,\lambda_1^2\,\psi_1+\frac{i}{2}\,u\,(2\,\lambda_1\,\phi_1-\bar{u}\,\psi_1)+{\psi_1}_t+\frac{i}{2}\,\phi_1\,u_x=0,
\end{equation}
\begin{equation}\label{non4}
\frac{i}{2}(2\,\lambda_1^2+u\,\bar{u})\,\phi_1-i\,\lambda_1\,\bar{u}\,\psi_1+{\phi_1}_t-\frac{i}{2}\,\psi_1\,\bar{u}_x=0.
\end{equation}

\begin{figure} 
		\includegraphics[width=8. cm,height=5.8 cm]{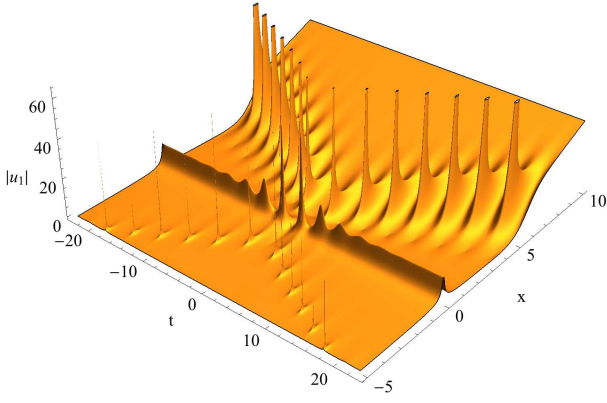}
	\caption{Solution (\ref{solnon}) of the nonlocal NLSE (\ref{nonNLSE}). It shows the interaction between a stationary solitons and two breathers, 
the breather soliton on the right side is in an inclined background and the left breather is in flat background. Parameters are:  $\lambda_{1i}=0.08$ and $\lambda_{1r}=1$. }
	\label{nonlocalfig}
\end{figure}

Applying  the seed solution, $ u_0=1/x$,  in (\ref{non1})-(\ref{non4}) and solving for $\psi_1$ and $\phi_1$, one obtains: 
\begin{equation}\label{non2}
\psi_1(x,t)=\frac{1}{x}\left[c_1-c_2\,(2\,\lambda_1\,x-1)\,e^{2\,\lambda_1\,(x+i\,\lambda_1\,t)}\right]e^{-\,\lambda_1\,(x+i\,\lambda_1\,t)},
\end{equation}
\begin{equation}\label{non3}
\phi_1(x,t)=\frac{1}{x}\left[c_1+c_2\,e^{2\,\lambda_1\,(x+i\,\lambda_1\,t)}+2\,c_1\,\lambda_1\,x\right]e^{-\,\lambda_1\,(x+i\,\lambda_1\,t)}.
\end{equation}
Then, the  new exact solution to the nonlocal NLSE (\ref{nonNLSE}) is obtained by substituting (\ref{non2}) and  (\ref{non3})  into (\ref{pot15})
\begin{equation} \label{non40}
u_1(x,t)=\frac{z_1(x,t)}{z_2(x,t)},
\end{equation}
where 
$z_1(x,t)=c_1^2\,(-2\,\lambda_2 \,x+e^{2\,\lambda_2\,(\,x+i\, \lambda_2\, t)})+\,c_1\, c_2 \,((4 \,\lambda_2^2 \,x^2 -1)\,e^{2\,\lambda_2\, (\,x+\,i \,\lambda_2\, t )}+(1-2\, \lambda_1\, x)\times\\e^{2\,x(\,\lambda_1 +\,\lambda_2)+2\, i\, t(\,\lambda_1*2+\,\lambda_2*2)} -2\, x(\,\lambda_2+2\,\lambda_1\, x(\,\lambda_1-\,\lambda_2))\, \times\\e^{2\,\lambda_1(\,x+\,i \,\lambda_1 t)}+2\,\lambda_2\, x+\,c_2^2\, (2\, \lambda_2\, x\,(1-2\, \lambda_1\, x))e^{2\, \lambda_1(x+i\, \lambda_1 t)}+(2\, \lambda_2 \,x-1)(4\,\lambda_1\, x^2\,(\,\lambda_1-\,\lambda_2)-2\, x\,(\,\lambda_1-\,\lambda_2)+1))$,\\
and \\
$z_2(x,t)=x\,(-2\,c_1^2 \,x\,(\,\lambda_1-\,c_2 \,\lambda_2\, e^{\,\lambda_2(2\, x+\,i\, \lambda_2 \,t)}+\,c_2^2\,(-1+2\, x \,\lambda_2)\,e^{2\,x\,(\,\lambda_1+\,\lambda_2)+2\, i\, t\, 
(\,\lambda_1^2 +\,\lambda_2^2)}-c_1\, c_2\,\times\\(2 \,x\,\lambda_1\, e^{2\,\lambda_1\, (\,x+\,i\, t\, \lambda_1)}+2\, c_2\, x \,\lambda_2 \,(-1+2\, x \,\lambda_1)\, e^{2\,x(\,\lambda_1+\,\lambda_2)+\,i\, t\, (2 \,\lambda_1^2+\,\lambda_2^2)}-(1+2\, x\, \lambda_1)(-1+2\, x\, \lambda_2)\, e^{2\, \lambda_2(\,x+\,i\, t \lambda_2)})))$, \\
where $c_1$ and $c_2$ are arbitrary real constants.
For simplicity we take $c_1=c_2=1$ thus, the solution $u_1(x,t)$ takes the following form:
\begin{widetext}
\begin{equation}\label{solnon}
u_1(x,t)=\frac{-4\,i\,{\lambda_1}_i(x\,|\lambda_1|^2-{\lambda_{1r}})e^{2\,q_3(x,t)}-{\lambda_1^*}^2e^{q_1(x,t)}+\lambda_1^2e^{q_2(x,t)}}{2\,i\left\{{\lambda_1}_i\,\text{cos}[q_4(x,t)]+{\lambda_1}_i\,\text{cosh}[2\,{\lambda_1}_r\,x-2\,i\,({{\lambda_1}_i}^2-{{\lambda_1}_r}^2)\,t]+|\lambda_1|^2\,x\,\text{sin}[q_4(x,t)]\right\}e^{q_3(x,t)}},
\end{equation}
\end{widetext}
where $q_1(x,t)=2\,\lambda_1^*\,(x+i\,\lambda_1^* t)$, $q_2(x,t)=2\,\lambda_1\,(x+i\,\lambda_1 t)$,\\
$q_3(x,t)=2\,x\,{\lambda_1}_r-2\,i\,({{\lambda_1}_i}^2-{{\lambda_1}_r}^2)\,t$, and $q_4(x,t)=2\,{\lambda_1}_i\,(x+2\,i\,{\lambda_1}_r t)$.\\
Solution (\ref{solnon}) corresponds to the scattering of a stationary soliton and two breathers; one on a flat background and the other is on an inclined background
as is shown in Fig. \ref{nonlocalfig}.

\section{New Exact Solutions to the Reverse-Time NLSE }\label{model3}
Another interesting possibility is an NLSE which is nonlocal in time rather than space, as for example 
\begin{equation} \label{reverNLSE}
i u_t+\frac{1}{2} u_{xx}-u^2 \tilde {u}=0,
\end{equation}
where $\tilde u=u^*(x,-t)$.
It was found that this equation admits a Lax pair only when the coefficient of the time derivative term is real \cite{Ab,Ab1}. 
Nevertheless, in the following discussion, we present three exact solutions to (\ref{reverNLSE}) using the traditional separation-of-variables method.\\
\begin{flushleft}
	\bf
	(i) $t-$independent solution: 
\end{flushleft}
Let us write 
\begin{equation}
u(x,t)=F(x),
\end{equation}
and  substitute into (\ref{reverNLSE}) to get
\begin{equation}\label{f}
F^{\prime\prime}(x)-2 F^3(x)=0,
\end{equation}
with the solution 
\begin{equation}
F(x)=\frac{c}{c\,x-1},
\end{equation}
where $c$ is an arbitrary real constant.\\ 
\begin{flushleft}
	\bf
	(ii) $x-$independent solution: 
\end{flushleft}
 We express the solution as 
\begin{equation}
u(x,t)=Z(T),
\end{equation}
where $T=i t$. 
Inserting in Eq. (\ref{reverNLSE})  leads to
\begin{equation}
Z^{\prime}(T)+ Z^3(T)=0,
\end{equation}  
with the solution 
\begin{equation}
u(x,t)=\pm\frac{ 1}{\sqrt{2it-2c}},
\end{equation}
where $c$ is an arbitrary real constant.\\
\begin{flushleft}
	\bf
	(iii) $t-$ and $x-$dependent solution: 
\end{flushleft}
 Finally, we express the solution as
\begin{equation}
u(x,t)=Z(T)\,e^{i\,x}.
\end{equation}
Substituting this into Eq. (\ref{reverNLSE}) leads to
\begin{equation}
Z^{\prime}(T)+ Z^3(T)+Z(T)=0,
\end{equation}  
which yields to the following exact solution to  (\ref{reverNLSE}) 
\begin{equation}
u(x,t)=\pm\frac{c\,e^{i\,x}}{\sqrt{e^{it}-2\,c^2}},
\end{equation}
where $c$ is an arbitrary real constant.  \\
Using any of these solutions as a seed for the DT will lead to higher order solutions corresponding to either a multi-singular soliton solution or breather.

\section{Conclusion} \label{Conc}

In this paper,  we derived a new two-soliton solution of the local NLSE  using the LP and  DT method  with an algebraically-decaying seed solution. 
Within this seed, the DT method leads to the so-called singular molecule soliton which is a higher-order solution composed of two diverging  peaks.  
The constructed solution allowed us to control practically all the characteristics of the molecule such as the binding energy, 
the force, and potential of interaction between the two solitons. 
We discussed in addition the scattering properties of such solitons.
The time evolution of the soliton width has  been also addressed to reveal the survival of these new structures. 
Furthermore, employing the LP in DT with the same seed solution as in the local NLSE case, 
we obtained  a new exact solution presented  an elastic interaction between one soliton and two breather solitons, on a flat and ramp backgrounds. 
The case with reverse-time nonlocal NLSE has been also highlighted.

While the solitons found here are singular and hence have a diverging norm, they may describe realistic situations such as the collapsing dynamics of a Bose-Einstein condensate 
when the repulsive interatomic interactions are switched to attractive \cite{PS}. 
In addition, a recent preprint by Sackakouji {\it et} al. \cite{Hid} has argued that such singular solitons may have some realistic relevance under certain circumstances.

 \section*{Acknowledgements}
K. M. Elhadj and A. Boudjem\^{a}a acknowledge support from the University of Chlef. L. Al Sakkaf and U. Al Khawaja acknowledge support 
from the  UAE University through the Grants N° UAEU-UPAR(4) and UAEU-UPAR(6). 
K. M. Elhadj is grateful to UAEU for hosting him during part of the work on this paper.

\end{document}